\documentclass[prb, onecolum, amsfonts]{revtex4}

\usepackage[dvips]{graphicx}
\usepackage{textcomp}
\usepackage{bm}

\begin{document}

\title{Phase reciprocity of spin-wave excitation by a microstrip antenna}

\author{T. Schneider}\email{tschneider@physik.uni-kl.de}
\author{A. A. Serga}
\author{T. Neumann}
\author{B. Hillebrands}
\affiliation{Fachbereich Physik, Technische Universit\"at Kaiserslautern, 67663 Kaiserslautern, Germany}
\author{M. P. Kostylev\footnote{On leave from St.Petersburg Electrotechnical University, St.Petersburg, Russia}}
\email{kostylev@cyllene.uwa.edu.au} \affiliation{School of Physics, University of Western Australia, Crawley, WA
6009, Australia}

\begin{abstract}
Using space-, time- and phase-resolved Brillouin light scattering
spectroscopy we investigate the difference in phase of the two
counterpropagating spin waves excited by the same microwave
microstrip transducer. These studies are performed both for backward
volume magnetostatic waves and magnetostatic surface waves in an
in-plane magnetized yttrium iron garnet film. The experiments show
that for the backward volume magnetostatic spin waves (which are
reciprocal and excited symmetrically in amplitude) there is a phase
difference of $\pi$ associated with the excitation process and thus
the phase symmetry is distorted. On the contrary, for the
magnetostatic surface spin waves (which are non-reciprocal and
unsymmetrical in amplitude) the phase symmetry is preserved (there
is no phase difference between the two waves associated with the
excitation). Theoretical analysis confirms this effect.

\end{abstract}

\maketitle

\section{Introduction}
It is well known that two very different kinds of dipolar dominated
spin waves can propagate in an in-plane magnetized magnetic film
\cite{D-E}. One of them, the so called backward volume magnetostatic
spin wave (BVMSW, magnetic bias field in the film plane and parallel
to the propagation direction), is characterized by a harmonic
distribution of the dynamic magnetization across the film thickness
and is completely reciprocal. The other one, the magnetostatic
surface spin wave (MSSW, magnetic bias field is in-plane, but
perpendicular to the propagation direction), has an exponential
distribution of the dynamic magnetization on the film thickness.
When the spin-wave propagation direction is reversed the maximum of
the dynamic magnetization of this mode shifts from one film surface
to the other. The surface wave is thus non-reciprocal. Influence of
the reciprocity is important for many spin-wave phenomena connected
both with parametric \cite{reversal-APL,reversal-JAP} and nonlinear
\cite{collision, TechPhys} interaction of counterpropagating waves,
including reflection \cite{Kovshikov} and excitation \cite{Tsankov}.

One of the most evident manifestations of this influence is that the
amplitudes of two counterpropagating spin waves excited by a
microwave current through an antenna mounted on top of the magnetic
film significantly depend on the used geometry. In the case of
reciprocal BVMSWs these waves propagate with identical amplitudes.
If one excites the non-reciprocal MSSWs this amplitude symmetry is
drastically distorted and spin-wave propagation in one direction is
strongly suppressed. This amplitude effect is immediately visible
and thus always taken into account in experiments on excitation of
spin waves by microstrip transducers placed near the film surfaces.

In contrast to the spin-wave amplitude the reciprocity of the phase
of spin waves excited that way is not obvious. In this article we
demonstrate that the phase symmetry is distorted when the reciprocal
BVMSW is excited and is preserved in the case of the non-reciprocal
MSSW. As a measure of the phase symmetry we used the difference in
initial phases of the two counterpropagating waves excited by the
same transducer. In the following we will term this phase difference
as the \lq\lq excitation phase\rq\rq.

The importance of the excitation phase is clear in all cases when
linear interference or nonlinear interaction of the excited
counterpropagating waves is possible. This is the usual situation
for rectangular and especially for ring resonators (see e.g.,
Refs.~\onlinecite{Kunz} and \onlinecite{Poston}). The influence of
the excitation phase has also been pointed out in the previous works
on spin wave soliton collision \cite{collision, TechPhys} and on
nonlinear pattern formation \cite{rectangle}. In particular, in
Ref.~\onlinecite{rectangle} the excitation phase determines the
selection rule for excitation of nonlinear standing-wave spin wave
resonances in a rectangular monodomain magnetic square, which is
important from the point of view of switching elements of magnetic
memory devices. Knowledge of the initial phase of spin waves is also
important for recently suggested spin wave logic applications
\cite{Logic-Kostylev, Logic-Schneider, Splitter}.

The conventional method of spin-wave detection using for example two
microstrip antennae equally spaced on both sides of the input
transducer cannot be used to measure the excitation phase due to the
symmetry of the excitation and detection process. Detecting the spin
waves using this technique would lead to an additional phase shift
of the same magnitude than the excitation phase but of the opposite
sign. The microwave signals in the output antennae will thus always
be in phase, regardless of the initial phase difference between the
two spin waves. This problem is general for all situations when
detection is performed via inducted microwave currents. It can now
be avoided due to the recent development of phase resolved Brillouin
light scattering (BLS) spectroscopy technique \cite{APL-Phase}, an
extension of the space- and time-resolved BLS \cite{Buett}, and the
possibility to observe variation of spin-wave phase with propagation
distance (the so called phase accumulation \cite{EPL-Phase}), which
allows us to measure the excitation phase of microwave excited spin
waves using a purely optical technique.

Using the described method we were able to directly observe the
difference in excitation phase between the BVMSW and MSSW geometry.
We also suggest a theoretical explanation for the observed effect.

\section{Results for the BVMSW configuration}

\begin{figure}
  \centering
  \includegraphics[width=\columnwidth]{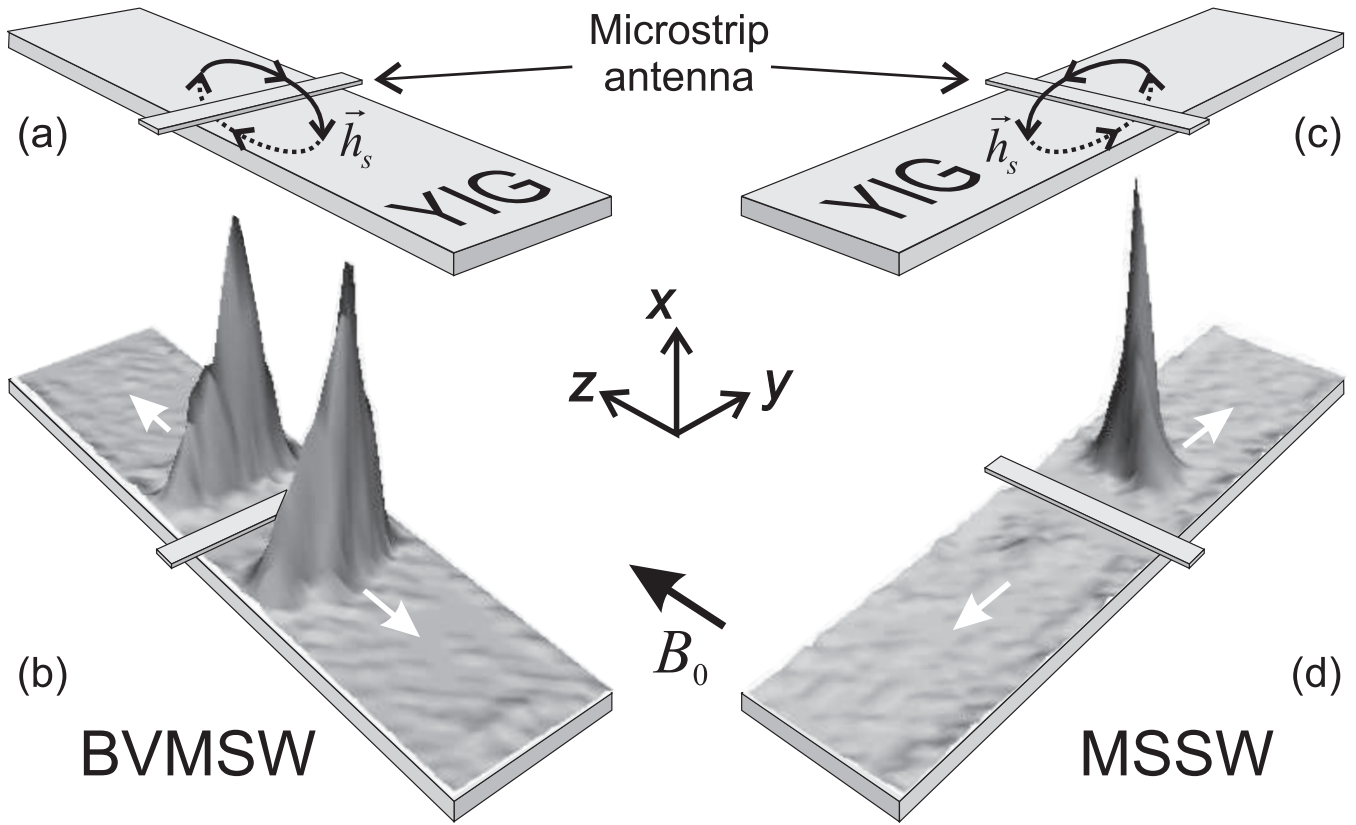}
  \caption{(a) Experimental setup for BVMSW geometry. Spin waves were
  excited using a microstrip antenna. (b) Space resolved BLS
  measurement of the intensity of two counterpropagating spin waves in
  the BVMSW geometry approximately 40\,ns after their excitation. (c)
  Experimental setup for MSSW geometry. (d) Space resolved BLS
  measurement of two counterpropagating spin waves in the MSSW
  geometry approximately 40\,ns after their excitation.}\label{Setup}
\end{figure}

Figure~\ref{Setup}(a) shows a picture of the experimental setup.
Spin waves are excited using a  microstrip antenna placed on top of
a 5\,\textmu m thick yttrium iron garnet (YIG) spin-wave waveguide.
The antenna width $w$ is 50\,\textmu m. Spin waves excited by
microwave pulses applied to the antenna can propagate freely in both
directions, as demonstrated in part (b) of the figure for BVMSW
geometry.

\begin{figure}
  \centering
  \includegraphics[width=\columnwidth]{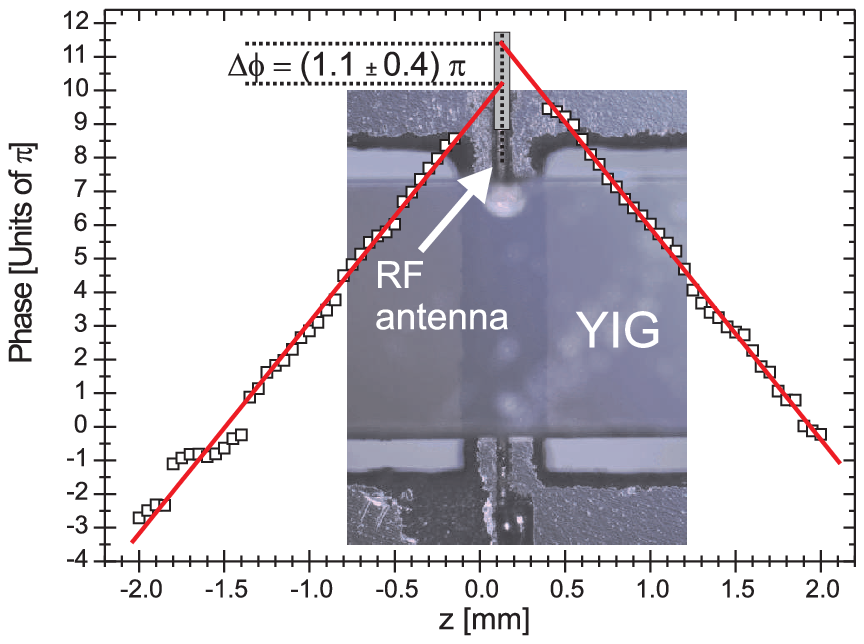}
  \caption{(Color online) Phase accumulation for BVMSW geometry. The
  measured excitation phase is $(1.1\pm0.4)\pi$. The applied bias
  magnetic field is 1835 Oe. Spin wave carrier frequency and
  wavevector are 7.132\ GHz and $(-198\pm1)\ \mathrm{cm}^{-1}$
  respectively. The inset shows a micrograph of the used
  structure.}\label{BVMSW-Exp}
\end{figure}

A typical example of the BVMSW phase accumulation measurement
performed in this work is shown in Fig.~\ref{BVMSW-Exp} (bias
magnetic field 1835\ Oe is applied in plane, parallel to the
propagation direction, as shown in Fig.~\ref{Setup}(a), carrier
frequency is 7.132\,GHz). One can clearly see a linear variation of
the phase $\phi(z)$. As we have used a linear, quasi monochromatic
wave, the accumulated phase can be described as wavevector $k$ times
the propagation distance $\left|z\right|$ (plus the initial phase).
One sees that the phase decreases with propagation distance as is
characteristic for a BVMSW mode.

Using a micrograph of the microstrip antenna and the substrate
holding it (shown in the inset of Fig.\ \ref{BVMSW-Exp}) one can
identify the position of the antenna in the measured data. The
extrapolation of the linear fit of the phase accumulation up to the
center of the antenna allows one to determine the excitation phase
of the spin wave under investigation as $(1.1\pm0.4)\pi$ in the
shown example. Other measurements performed for different bias
magnetic fields (and thus different wavevectors) confirm a value of
$\pi$ for the excitation phase of BVMSW within the error margins.

This result can be understood based on the configuration of the
external magnetic field and the distribution of the dynamic
magnetization associated with BVMSW excitation and propagation.
Using the coordinate system defined in Fig.\ \ref{Setup}(a) one
finds that the dynamic magnetization $\bm{m}$ lays in the
$x$--$y$--plane, while the microwave field created by the antenna
$\bm{h}_s$ has components in $x$-- and $z$--directions only. The
relation between the components of the effective dynamic field and
the dynamic magnetization is given by the microwave susceptibility
tensor $\hat\chi$ (see e.g. Ref.\ \onlinecite{Gurevich}). This
relation involves only the components perpendicular to the direction
of the vector of static magnetization. In our case the components of
the effective dynamic magnetic field which exert considerable
torques onto the magnetization vector are the antenna microwave
field $\bm{h}_s$ and the spin wave dipole field $\bm{h}_d$.
Therefore this relation takes the form
\begin{equation}
  \bm{m} = \hat\chi \cdot \left(\bm{h}_s + \bm{h}_d \right) \,.\label{m}
\end{equation}

In the following we will consider the dynamic magnetization and the
effective magnetic field averaged across the film thickness. In this
approximation the dipole field of long-wavelength spin waves is well
described by the quasi-one-dimensional tensorial Green's function
$\hat{G}$ (see Ref.\ \onlinecite{Guslienko}). This description is
valid for  $kL < 1.5$ (where $k$ is the spin wave wavevector and $L$
is the film thickness)\cite{Kostylev}. The function has only two
nonvanishing components, both of which are diagonal: the
{out-of-plane} $\rightarrow$ {out-of-plane} component $G^{(oo)}$ and
the {in-plane} $\rightarrow$ {in-plane} component $G^{(ii)}$. The
dipolar field $\bm{h}_d$ can be written as
\begin{equation}
  \bm{h}_d(r) = \int^\infty_{-\infty} \hat{G}(r-r')\bm{m}(r')dr' \,,\label{h}
\end{equation}
where $r$ is the coordinate along the spin wave propagation path. It
is important that both components are symmetric:
$\hat{G}(-s)=\hat{G}(s)$. Therefore the dipole field does not
introduce nonreciprocity into the propagation of long-wavelength
spin waves.

As BVMSWs possess only an out of plane component of the dipolar
field Eq.~(\ref{h}) can be written as
\begin{equation}
  h_{dx}=\int^\infty_{-\infty} G^{(oo)}(r-r') m_x(r')dr' \,.
\end{equation}

Furthermore, only the out-of-plane component of the antenna
microwave field ($h_{sx}$) lies in the plane perpendicular to the
static magnetization. Under these conditions Eqs.\ (\ref{m}) and
(\ref{h}) reduce to an inhomogeneous scalar integral equation
derived in Ref.\ \onlinecite{scattering} that involves only $m_x$,
$G^{(oo)}$, $h_{sx}$, and the diagonal component of $\hat\chi$. This
results in equal amplitudes for the waves propagating in positive
and negative direction. It also translates the symmetry of
$h_{sx}(z)$ to the phase accumulation profile $\phi(z)$. Since
$h_{sx}(z)$ has an antisymmetric profile along the z-axis with a
node in the middle of the antenna this leads to an excitation phase
of $\pi$.

Indeed, as follows from Eq.\ (B18) in Ref.\ \onlinecite{scattering},
far away from the antenna ($|z| \gg w/2$) the expression for the
dynamic magnetization takes the form
\begin{equation}\label{BVm}
  m_x(\pm \left|z\right|)=\frac{4\pi i}{L} h_{\mp\left|k\right|sx}
  \exp\left(i\left|kz\right|\right)\,.
\end{equation}

Note, that the sign in front of the complex exponent is positive
since the waves are backward ones. This results in the observed
phase accumulation $\phi (y)=-\left|k z\right|$\,. The magnitude
$h_{k sx}$ is the spatial Fourier component of $h_{sx}(z)$
corresponding to the spatial frequency $k$. As mentioned before
$h_{sx}(z)$ is an odd function of $z$ with a node in the middle of
the antenna, so $h_{-\mid k \mid sx}=-h_{\mid k \mid sx}$, which
gives the excitation phase of $\pi$ as confirmed in our
measurements.

\section{Results for the MSSW configuration}

\begin{figure}
  \centering
  \includegraphics[width=\columnwidth]{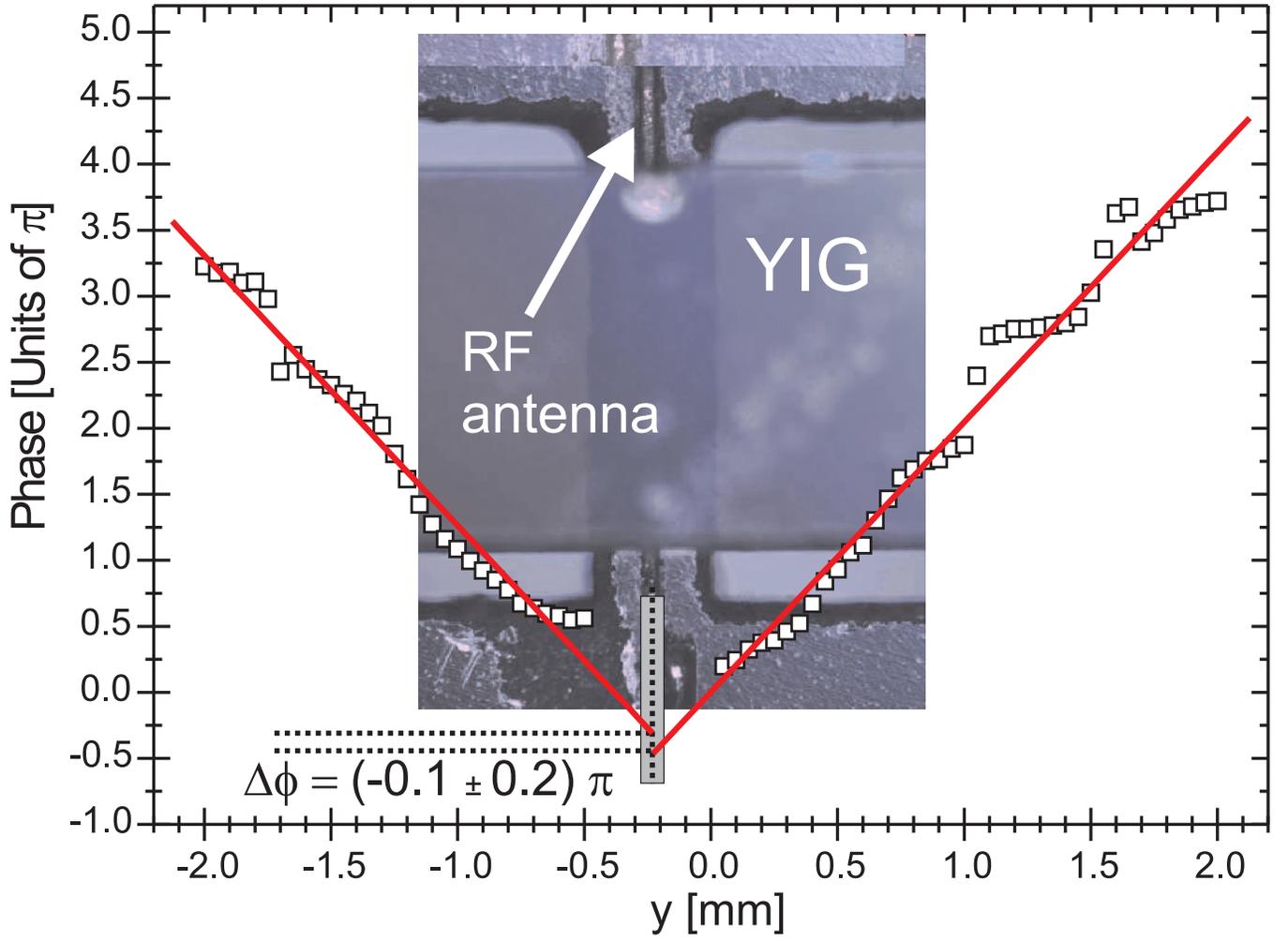}
  \caption{(Color online) Phase accumulation for MSSW geometry. The measured excitation
  phase is $(-0.1\pm0.2)\pi$. The applied bias magnetic field is
  1825 Oe. Spin wave carrier frequency and wavevector are 7.132\ GHz
  and $(64.21\pm0.04)\ \mathrm{cm}^{-1}$ respectively. The inset shows a
  micrograph of the used structure.}\label{MSSW-Exp}
\end{figure}

Figure\ \ref{MSSW-Exp} shows a typical example of the measurements
for MSSW geometry (bias magnetic field 1825\,Oe is applied in plane,
perpendicular to the propagation direction, as shown in Fig.\
\ref{Setup}(c), carrier frequency is 7.132\,GHz). As in the BVMSW
case the phase accumulation $\phi(y)$ on both sides from the antenna
is linear, but this time the phase increases with the propagation
distance. The excitation phase is determined as $(-0.1 \pm 0.2)\pi$.
Additional measurements for different bias magnetic fields confirm
the zero value of MSSW excitation phase within the error margins.

Thus in the MSSW geometry the situation is quite different from the
BVMSW case. This can be explained based on Eq.\ (\ref{m}). As it is
demonstrated in Fig.\ \ref{Setup}(c) for the MSSW case both fields
$\bm{h}_d$ and $\bm{h}_s$ as well as the dynamic magnetization
$\bm{m}$ lay in the $x$--$y$--plane. Therefore all components of the
vectors $\bm{h}_d$ and $\bm{h}_s$ in Eq.\ (\ref{m}) have to be taken
into account. Contributions of all these components are efficiently
intermixed by the microwave susceptibility tensor through the
products $\hat\chi\cdot\bm{h}_s$ and $\hat\chi\cdot\bm{h}_d$. The
result of intermixing is not trivial and can be space and wavevector
dependent, because: (i) the anti-diagonal components of the
microwave susceptibility tensor are shifted in phase by $\pm \pi /2$
with respect to the diagonal components; (ii) below the film surface
the in-plane component of the antenna field $h_{sy}$ has the same
phase everywhere, whereas the out-of-plane component $h_{sx}$
changes its phase by $\pi$ at the longitudinal antenna axis $y=0$;
and (iii) the in-plane and the out-of-plane components of the
Green's function are dependent on the spin-wave wavevector in
different ways (see e.g., Ref.\ \onlinecite{guided modes}).

Thus the behaviour of the complex amplitude of MSSWs can be
nontrivial. A well-known manifestation of this behaviour is the
nonreciprocity of MSSW excitation by microstrip transducers. These
waves propagate with significantly different amplitudes in the
positive and negative directions from a microstrip antenna. This
MSSW nonreciprocity is clearly seen in our BLS measurement shown in
Fig.\ \ref{Setup}(d).

The system of equations\ (\ref{m}) and (\ref{h}) for MSSWs allows
for an analytical solution. As in the case of BVMSWs
\cite{scattering} the solution involves a spatial Fourier
transformation of the Green's function of the dipole field. An
explicit formula relating the complex vector $\bm{m}$ to $\bm{h}_s$
can be found e.g., in Ref.\ \onlinecite{Kalinikos}. Here we obtain a
solution in a form which allows for an expression for the excitation
phase. Following the same method as in Ref.\ \onlinecite{scattering}
from Eqs.\ (\ref{m}) and (\ref{h}) one arrives at the solution for
the dynamic magnetization vector
\begin{equation}
  \bm{m}(y)=\int^{+\infty}_{-\infty}\hat
  G_\mathrm{exc}(y-y')\bm{h}_s(y')dy'\,,\label{DEm}
\end{equation}
where
\begin{equation}
\hat G_\mathrm{exc}(y-y')=\frac{1}{D(k)} \hat Y F(y-y')\,,\label{Green}
\end{equation}
and the matrix $\hat Y$ has the elements
\begin{eqnarray}
  Y_{11}&=&-W(k)-1+\frac{\chi}{(\chi_a^2-\chi^2)}\nonumber\\
  Y_{12}&=&\frac{i\chi_a}{(\chi_a^2-\chi^2)}\nonumber\\
  Y_{21}&=&-\frac{i\chi_a}{(\chi_a^2-\chi^2)}\nonumber\\
  Y_{22}&=&W(k)+\frac{\chi}{(\chi_a^2-\chi^2)}\,,\label{Y}
\end{eqnarray}
and
\begin{eqnarray}
F(s)&=&\frac{1}{2 \pi}\left[-2 \pi i \ \exp(-i k_c | s |)\right.\nonumber\\
&&+\exp(-i k_c^* | s |) E_1(-i k_c^* | s |)\nonumber\\
&&+\exp(i k_c^* | s |)\left.\! E_1(i k_c^* | s |)\right]\,,\label{F}
\end{eqnarray}
where the asterisk ($^*$) denotes  complex conjugation, and $E_1(z)$
is the exponential integral \cite{Abramovitz}. In these expressions
the complex wavenumber $k_c$ is defined as $k_c=k+i\nu$, where $k>0$
represents the carrier wavenumber of the excited MSSW. It can be
calculated as the root of the approximate dispersion relation
\begin{equation}
W(k)\left(W(k)+1\right)+ \mathrm{Re}\left[\frac{\chi+1}{\chi_a^2-\chi^2}\right]=0\,,\label{dispersion}
\end{equation}
where $W(k)=\left(\exp(-\left|k\right|L)-1\right)/(\left|k\right|L)$
and $\chi$ and $\chi_a$ are the diagonal and off-diagonal components
of the microwave magnetic susceptibility tensor $\hat\chi$ for the
film material \cite{Gurevich}. The tensor components are complex
numbers, as magnetic losses in the material are included.

The imaginary part of $k_c$ determines the spatial spin wave damping
and is negative ($\nu \leq 0 $) for MSSWs:
\begin{equation}
  \nu=-\left(D(k)\right)^{-1}\mathrm{Im}\left[\frac{\chi+1}{\chi_a^2-\chi^2}\right]\,,
\end{equation}
where
\begin{equation}
  D(k)=\frac{\mathrm{d}}{\mathrm{d}(\left|k\right|)}\left(W(k)(W(k)+1)\right)<0\,.
\end{equation}
The negative value of $\nu$ shows that the direction in which the
wave amplitude decreases due to damping coincides with the direction
in which the wave phase $\phi(y)$ increases (see Eq.\ (\ref{F})).
The wave is therefore a forward one which is consistent for MSSWs
(for comparison: for BMSWs $\nu>0$ and the waves are backward ones
\cite{scattering}).

As one sees from Eqs.\ (\ref{DEm})-(\ref{F}), the scalar function
$F(s)$ determines the phase accumulation while the matrix-vector
product $\frac{1}{D(k)}\hat Y\cdot\bm{h}_s $ determines the MSSW
initial phase and hence is responsible for the excitation phase.
Indeed, like in the BVMSW case \cite{scattering}, the near
(unretarded) field of the antenna is proportional to the terms in
(\ref{F}) involving the exponential integrals $E_1(s)$ the sum of
which is real. Far away from the antenna the near  field vanishes
and the $y$-dependence of the dynamic magnetization is determined by
the first term in the brackets in Eq.\ (\ref{F}) which is an
intrinsically complex function of $s$ and is purely imaginary at the
source point $s=0$. This term describes a retarded field (traveling
waves) propagating in both directions from the excitation source:
\begin{equation}
  F(s)=-i\exp(-i\left|k s\right|)\,.\label{F2}
\end{equation}
(In this formula and onwards for the sake of simplicity we ignore
magnetic losses by setting $\nu=0$.) In contrast to Eq.\ (\ref{BVm})
the sign of the exponent is negative. This results in a positive
slope for the phase accumulation $\phi (y)=\left|ky\right|$ for
MSSWs as demonstrated by our experiments.

All the expressions above are valid for long-wavelength MSSW with
carrier wavevectors $k$ for which $W(k)<-0.5$\,. For very small
wavevectors ($kL \ll 1$) one can use the approximation $W(k)\approx
-1+(\left|k\right|L)/2$\,, $D(k)\approx L/2=D$\,.  Furthermore, for
$\nu=0$ $\chi$ and $\chi_a$ are real and for $kL \ll 1$ from the
dispersion relation Eq.\ (\ref{dispersion}) one obtains
$k=2(\chi+1)/[L(\chi^2_a-\chi^2)]$. Then the elements of the matrix
$\hat S=\hat Y/D$ reduce to
\begin{eqnarray}
  S_{11}&=&-\frac{2}{L(\chi_a^2-\chi^2)}\nonumber\\
  S_{12}&=&\frac{2i\chi_a}{L(\chi_a^2-\chi^2)}\nonumber\\
  S_{21}&=&-\frac{2i\chi_a}{L(\chi_a^2-\chi^2)}\nonumber\\
  S_{22}&=&\frac{2[(\chi+1)^2-\chi_a^2]}{L(\chi_a^2-\chi^2)}\,.\label{S}
\end{eqnarray}

Using Eqs.\ (\ref{Green}), (\ref{F2}), and (\ref{S}), Eq.\
(\ref{DEm}) is then transformed into
\begin{equation}
  \bm{m}\left(\pm\left|y\right|\right)= -{2\pi i \hat S} \exp(-i\left|k y\right|)\bm{h}_{\pm\left|k\right|s}\,,
  \label{14}
\end{equation}
where $\bm{h}_{ks}=(2\pi)^{-1}\int_{-\infty}^{\infty}\bm{h}_{s}(y)
\exp(iky)dy$ is the spatial Fourier transform of the antenna field.
From this formula one sees that the initial phases of the excited
waves are determined by the phase of the product $\hat S
\bm{h}_{\pm\left|k\right|s}$.

As discussed in the previous section, the Fourier transform of the
out-of-plane component of the antenna field is an odd function of
$k$. On the contrary, since the spatial profile of the in-plane
component $h_{sy}(y)$ is symmetric with respect of the longitudinal
antenna axis $y=0$, its Fourier transform is an even function and
thus $h_{-\left|k\right|sy}=h_{\left|k\right|sy}.$ Furthermore, from
Maxwell equations one finds that for a microwave field generated by
a current through a microstrip transducer it stands $h_{ksx}=i\,
\mathrm{sign}(k)h_{ksy}$ (see e.g., Eq.\ (32) in Ref.\
\onlinecite{Dmitriev}). Accounting for the change in sign in Eq.\
(\ref{14}) this formula can be written as
$h_{ksx}=i\,\mathrm{sign}(y)h_{ksy}$ and the vector
$\bm{h}_{\pm\left|k\right|s}$ in Eq.\ (\ref{14}) takes the form
\begin{equation}
  \bm{h}_{\pm\left|k\right|s}=\left(\begin{array}{c} i\, \mathrm{sign}(y)\\
  1\end{array}\right)h_{\left|k\right|sy}\, .\label{15}
\end{equation}
One now sees that the MSSW initial phase is determined by the
product of the matrix $\hat S$ with the vector
$(i\,\mathrm{sign}(y), 1)^\mathrm{T}$.

It is sufficient to consider only the first element of the resulting
vector (the second one differs from the first one only by the
coefficient of ellipticity of the precession of the magnetization
vector and by the phase shift of $\pi /2$). Using Eqs.\ (\ref{14})
and (\ref{15}) $m_{x}$ can be written as
\begin{equation}
  m_{x}(\pm\left|y\right|)=\frac{4\pi}{L}h_{\left|k\right|sy}\frac{\chi_a-\mathrm{sign}(y)}{\chi_a^2-\chi^2}\exp(-i\left|ky\right|).\label{16}
\end{equation}

As in our range of frequencies and static fields $\chi_a<-1$, the
real coefficient in front of the complex exponent in Eq.\ (\ref{16})
has the same sign for positive and negative $y$-values, i.e., for
the positive and the negative waves excited by the same antenna.
This shows that the MSSW excitation phase is zero, as confirmed by
our measurements.

\section{Summary and conclusion}

\begin{figure}
  \centering
  \includegraphics[width=\columnwidth]{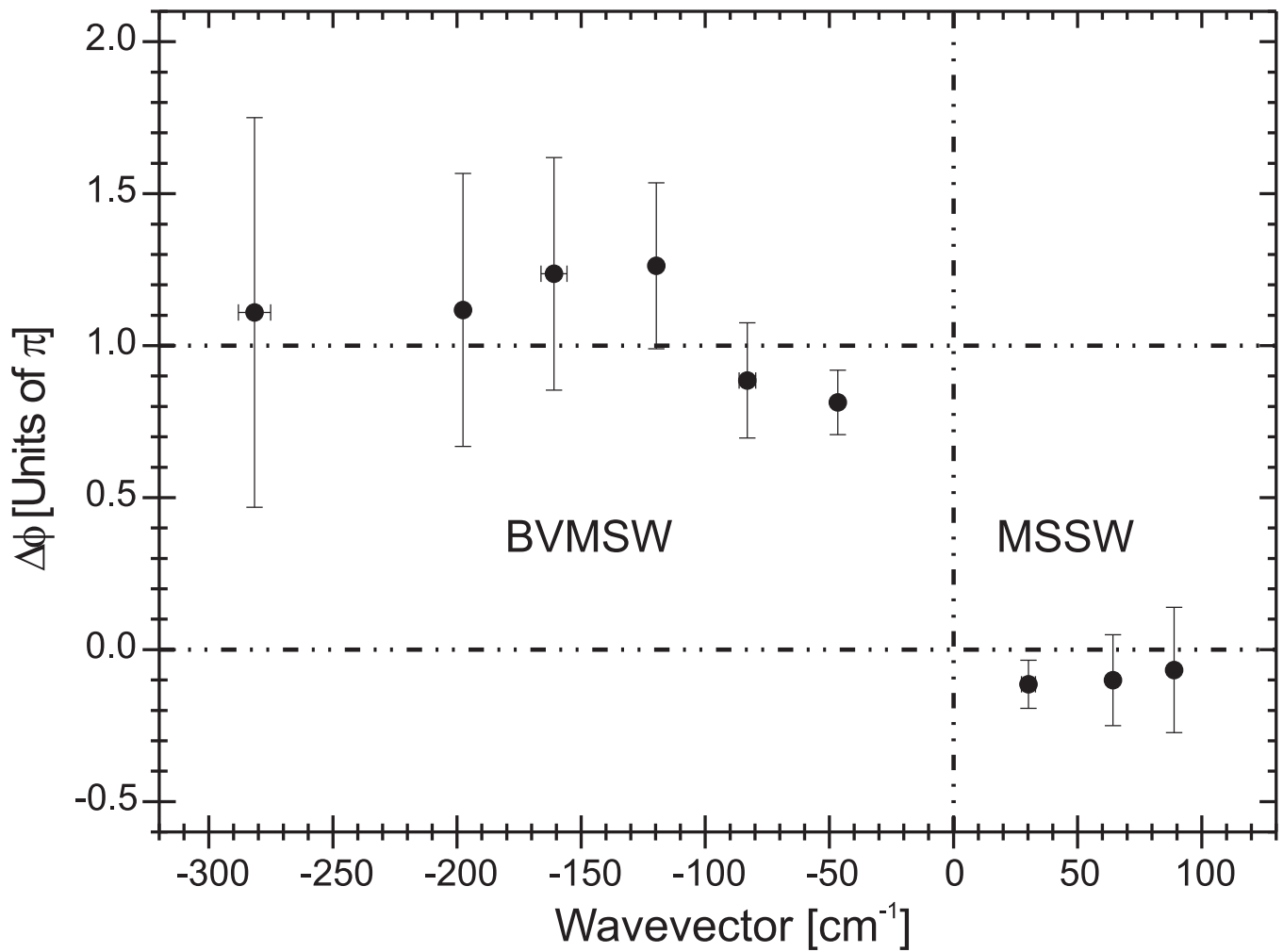}
  \caption{Summary of the performed measurements. The expected
  behaviour is clearly visible. The different excitation phase
  values for the lowest absolute values of the wavevectors can be
  explained by the distortion due to the propagation close to the
  ferromagnetic resonance frequency.}\label{Summary}
\end{figure}

Figure \ref{Summary} shows a summary of all performed experiments.
The above-described behaviour is clearly visible for all
wavevectors. The notably different excitation phase for the
measurements with the lowest absolute value for the wavevector can
be explained by taking into account that a spin-wave pulse
propagating close to the ferromagnetic resonance frequency becomes
strongly distorted. This distortion also influences the phase and
thus the measured values for the excitation phase. Note that the
main source of error is the position of the antenna. Since this
position error translates to the phase error as $k\cdot\Delta x$ the
error margins become larger for larger wavevectors.

In conclusion we have demonstrated that two counterpropagating
backward volume waves which are excited by the microstrip transducer
symmetrically in amplitude show an excitation phase of $\pi$ while
the non-reciprocal surface waves which are unsymmetrical in
amplitude have zero excitation phase.

\begin{acknowledgments}
Financial support by the Deutsche Forschungsgemeinschaft
(Graduiertenkolleg 792 and MATCOR) and the Australian Research
Council and technical support from the Nano + Bio Center, TU
Kaiserslautern is acknowledged.
\end{acknowledgments}

\bibliography{}

\end{document}